# Development of the MICROMEGAS Detector for Measuring the Energy Spectrum of Alpha Particles by using a $^{241}$Am Source

Do Yoon Kim and Cheolmin Ham

*Department of Energy Science, Sungkyunkwan University, Suwon 440-746, Republic of Korea*

Jae Won Shin, Tae-Sun Park and Seung-Woo Hong

*Department of Physics, Sungkyunkwan University, Suwon 440-746, Republic of Korea*

Samuel Andriamonje and Yacine Kadi

*CERN CH-1211, Geneva, Switzerland*

Claudio Tenreiro

*Faculty of Engineering, University of Talca, Talca 3465548, Chile*

We have developed MICROMEGAS (MICRO MEsh GASeous) detectors for detecting α particles emitted from an $^{241}$Am standard source. The voltage applied to the ionization region of the detector is optimized for stable operation at room temperature and atmospheric pressure. The energy of α particles from the $^{241}$Am source can be varied by changing the flight path of the α particle from the $^{241}$Am source. The channel numbers of the experimentally-measured pulse peak positions for different energies of the α particles are associated with the energies deposited by the alpha particles in the ionization region of the detector as calculated by using GEANT4 simulations; thus, the energy calibration of the MICROMEGAS detector for α particles is done. For the energy calibration, the thickness of the ionization region is adjusted so that α particles may completely stop in the ionization region and their kinetic energies are fully deposited in the region. The efficiency of our MICROMEGAS detector for α particles under the present conditions is found to be ~ 97.3 %.

Email: swhong@skku.ac.kr

Fax: +82-31-290-7047

# I. INTRODUCTION

The MICROMEGAS (MICRO-Mesh-Gaseous Structure) detector [1, 2] is a type of micro-pattern gas detector [3-5]. The advantages of the MICROMEGAS detector include the ability of selective detection of neutral [6-8] and charged [9] particles with a relatively high spatial resolution [10, 11] and its operation at room temperatures with low gamma-ray sensibilities [12, 13]. It can be fabricated in a large size [14, 15] at reasonable costs and can be easily implemented in many areas such as nuclear and high-energy physics [16-18] and medical applications [19]. It has flexibility in the choice of the top cathode plate material and the gas mixture in the chamber, which can be exploited to produce optimal performance for selective detection of different particles. For instance, the detection of neutrons by using a MICROMEGAS detector can be done through the nuclear reaction $^{10}$B(n,α)$^{7}$Li with $^{10}$B deposited on the cathode plate [6-8], where the detection of α particles (and/or $^{7}$Li) is crucial. In this work, we have, thus, developed a MICROMEGAS detector for detecting α particles by using $^{241}$Am as a calibration source. For that, we need an energy calibration of our MICROMEGAS detector for α particles, which can be done by measuring α particles of various energies and comparing the results with those of a GEANT4 simulation.

GEANT4 simulations were done to calculate the energies of the α particles deposited in different regions of the MICROMEGAS detector, which enabled us to calibrate the detector, as will be described in Subsection III.3. As a source of α particles with variable energies, we use those from a standard $^{241}$Am source, and we use a gas as a degrader to change the energy. We find that the energy calibration can be done very well and that the detection efficiency of the detector is (97.3 ± 2.5)%.

# II. EXPERIMENTAL SETUP

1. Description of the Experimental Setup

Figure 1(a) shows our experimental setup, and Fig. 1(b) its schematic diagram. The MICROMEGAS detector is a double-stage parallel-plate gaseous detector, whose schematic diagram is shown by the dashed squares in Figs. 1(b) and 1(c). A $^{241}$Am source is placed on the top of the collimator, which is located on the MICROMEGAS detector. The gas volume of the MICROMEGAS detector is divided into two regions by a thin metallic micromesh represented by the thick dashed line in Fig. 1(c).

The upper part of the gas volume shown in Fig. 1(c) is referred to as the ionization region with a variable thickness $D_I$, in which α particles ionize the gas and create electrons, the latter of which are then attracted to the anode in the bottom. The lower part of Fig. 1(c) is referred to as the amplification region with a fixed thickness $D_M$ of 50 μm. The high electric field in the amplification region between the micromesh and the anode causes an avalanche; thus, the number of electrons is amplified in this region.

The region from just below the $^{241}$Am source to just above the cathode plate made of an aluminized mylar shall be referred to as the collimator region with the length of $D_C$ as shown in Fig. 1(b). If the collimator shown in Fig. 1(b) is thick, α particles need to travel a relatively long distance before they reach the cathode plate denoted by a thick horizontal line and enter the ionization region with thickness $D_I$. By changing the thickness of the collimator, we can vary the energies of the α particles from the $^{241}$Am source. Therefore, the collimator is used here not only to collimate the α particles but also to degrade the energies of the α particles to have variable energies. Just below the collimator is a cathode holder, which holds the cathode plate made of aluminized mylar. Below the cathode plate is the ionization region. The alphas that pass through the hole in the collimator and penetrate the cathode plate travel through the ionization region, ionize the gas, and generate electrons and ions.

The gas in the detector was chosen as a mixture of Ar 88% + $CF_4$ 10% + iso-$C_4H_{10}$ 2% [9, 11, 20] through GARFIELD simulation studies [21]. The generated electrons are accelerated toward the anode through the thin micromesh. In the amplification region which has the thickness $D_M$ of 50 μm, the electric field is strong enough to give rise to an avalanche, increasing the number of electrons exponentially. The amplified electrons are collected in the anode plate.

By changing the length of the collimator region from $D_C$ = 5.7 to 31.7 mm in the experiment, we varied the energy of the α particle entering the ionization region, which will be measured by using an 8 K multi-channel analyzer (MCA). We also varied the length $D_I$ of the ionization region from $D_I$ = 4 to 26 mm for measuring the energies of the α particles. If $D_I$ is not long enough, α particles may not stop in the ionization region and may leave the MICROMEGAS detector with part of their energy carried away. If $D_I$ is long enough, an α particle will stop in the ionization region and will fully deposit its kinetic energy in the ionization region, which we can use to calibrate the MICROMEGAS detector.

The data acquisition was done by using an ORTEC ASPEC-927 multi-channel analyzer and the Maestro-32 program. The read-out electronics consists of a CAEN N471A power supply, an ORTEC 142PC pre-amplifier, an ORTEC 672 amplifier, and a Tektronix digital oscilloscope TPS2024.

2. $^{241}$Am Source and Collimators

$^{241}$Am decays to $^{237}$Np, emitting an α particle with energies of ~ 5.485 MeV (~ 85 %), ~ 5.443 MeV (~ 13 %) and 5.388 MeV (~ 0.016 %) [22]. We used an $^{241}$Am source (3.585 kBq) purchased from Eckert & Ziegler, USA (item number EAB-241-PL), which had the shape of a flat circular dish with a rim. The diameter of the active area of the source is 45 mm, as shown in Fig. 2(b). α particles are emitted in random directions from the surface of the active area. Because α particles are randomly emitted from the $^{241}$Am source as illustrated in Fig. 4 of Ref. 23, we install collimators to collimate them. The collimator is made of polyethylene and has a hole of diameter of 2 mm in the center, as shown in Fig. 2(a). Figure 6 of Ref. 23 shows that the energies of the α particles from the $^{241}$Am source decrease as the flight path from the source gets longer in air. In Ref. 24, the installation of a collimator with a thickness of 6 mm is also shown to be able to induce a certain angular distribution in the fluence of the α particles from the $^{241}$Am source. By using this feature, we were able to vary the energies of the α from the source by installing collimators of different thicknesses.

The distance from the active surface of the $^{241}$Am source to the upper surface of the collimator is given by the height of the rim of the source (2.7 mm), as shown in Fig. 1(b). If we use a collimator with a thickness of 2 mm, the total distance ($D_C$) of the collimator region from the source to the cathode plate is a sum (5.7 mm) of the height of the source rim (2.7 mm), the thickness of the collimator material (2 mm), and the thickness of the cathode holder (1 mm), as shown in Fig. 1(b). We shall change the thickness of the collimator material from 2 to 28 mm, so that $D_C$ changes from 5.7 to 31.7 mm.

3. Simulation Tools

GEANT4 [25, 26] simulations were performed on the propagation of α particles from the $^{241}$Am source to various regions of the detector to estimate the energies of the α particles in different regions of the detector. In this work, GEANT4 version 10.00.p02 was used. The energy bin for scoring an α particle was taken as 0.02 MeV. Figure 3 shows the geometry and a snapshot of the propagation of an α particle from the $^{241}$Am source through the collimator hole. Most α particles are stopped by the collimator, as represented by the dots in Fig. 3. The solid lines represent the tracks of α particles released from the surface of the $^{241}$Am source and passing through a collimator hole of 2-mm diameter.

### III. RESULTS

1. Optimization of the Applied Voltages

We first optimized the voltages applied to the MICROMEGAS cathode, micromesh, and anode. In the optimization procedure, we used a 2-mm-thick collimator (thus, $D_C$ = 5.7 mm) with a fixed electric field $E_I = V_I/D_I = 0.25$ kV/cm (with $V_I$ = 100 V and $D_I$ = 4 mm) for the ionization region and let the voltage difference ($V_M$) on the micromesh vary from -170 to -280 V. (See Fig. 1 (c) for $V_I$ and $V_M$.) The amplitude of the output voltage was measured while checking the peak position with a MCA. The measured output voltage is plotted in Fig. 4(a) as a function of the electric field strength $E_M = V_M/50$ μm in the amplification region. As $E_M$ increases from 26 to 56 kV/cm (or $V_M = -130 \sim -280$ V), the amplitude of the output voltage increases exponentially for all different values of $D_I$, as shown in Fig. 4(a).

The same output voltage is plotted in Fig. 4(b) as a function of $D_I$ for various values of $E_M$. The output voltage is seen to increase linearly as a function of $D_I$ from 4 to 20 mm. Therefore, the output voltage can be nicely fitted by using the following expression as a function of $D_I$ and $E_M$:

$$V = E_{I0} \cdot D_I \cdot e^{E_M/E_{M0}}, \tag{1}$$

where the constants $E_{I0} = (1.31 \pm 0.08)$ V/cm and $E_{M0} = (7.24 \pm 0.08)$ kV/cm are independent of $E_M$ and $D_I$. Equation (1) shows that while the output voltage increases exponentially with respect to the electric field $E_M$ in the amplification region, it increases linearly with respect to the length of the ionization region $D_I$. This linear dependence of V on $D_I$ is, however, valid only for small values of $D_I$. Once $D_I$ becomes large enough to completely stop the incident alpha, the linear dependence will no longer hold.

Next, we fixed the electric field in the amplification region as $E_M = 36$ kV/cm, which is around the middle of the range of 26 kV/cm < $E_M$ < 56 kV/cm (Fig. 4(a)), varied the voltage difference ($V_I$) in the ionization region from -5 to -820 V, and measured the output voltage. Figure 4(c) shows a plateau of the output voltage for various lengths ($D_I$) of the ionization region. This plateau corresponds to the maximum generation of primary electrons in the ionization region and maximum transmission through the micromesh [27]. The electric fields corresponding to the plateau for various values of $D_I$ are given in Table 1. The optimal range of electric fields common to all the lengths of $D_I$ from 4 to 20 mm is around $E_I = 0.18 \sim 0.35$ kV/cm. We, thus, chose the electric field for the ionization region as $E_I = 0.18 \sim 0.35$ kV/cm and the electric field for the amplification region as $E_M = 36$ kV/cm for the experiments to be discussed henceforth.

2. Simulations

For the energy calibration of our MICROMEGAS detector, we need to know how much energy of the incident alpha is deposited in the 'ionization' region, because that is what is measured experimentally. To estimate the energy deposited by the α particles in the ionization region, we performed GEANT4 simulations for different thicknesses of the collimator. By comparing the energy deposit estimated by the GEANT4 simulations with the experimentally measured channel numbers for the pulse peak positions, we were able to do an energy calibration.

In the estimate of the energy deposited by the α particles in the ionization region, we need to know the kinetic energy of the α particles and the energy deposited by the α particles in each component of our MICROMEGAS detector. Let us denote the upper surface of the cathode plate by $S_1$ and the lower surface of the cathode by $S_2$, as shown in Figs. 1(b) and 1(c). The α particles from the $^{241}$Am source arrive at the upper surface $S_1$ and lose part of their energy in the cathode plate made of a 12-μm-thick mylar plate with 100 nm of aluminum deposited on the mylar. After losing part of its energy in the cathode plate, an α particle leaves the aluminized mylar cathode through the surface $S_2$. Figure 5(a) shows the calculated energies ($E_α$) of the α particles entering $S_1$ (denoted by the empty circles) and leaving $S_2$ (denoted by the empty diamonds) for various thicknesses ($D_C$) of the collimator region ranging from 5.7 to 27.7 mm. (Note that the maximum value of $D_C$ was chosen as 31.7 mm for the experiments as was mentioned in Subsection II.1 and will be discussed in Subsection III.3, but that of $D_C$ for simulations was 27.7 mm.) The (red) horizontal dotted line denotes the initial alpha energy of 5.485 MeV as mentioned in Section II.2. Figure 5(a) shows that $E_α$ at $S_1$ and $S_2$ decreases as $D_C$ increases, which shows that the gas in the collimator plays the role of a degrader. $E_α$ at $S_1$ is roughly ~ 2 MeV higher than that at $S_2$; thus, about 2 MeV is lost inside the cathode plate.

Let us denote the energy loss of an α particle in the collimator region by $\Delta E_C$, that in the cathode plate by $\Delta E_P$, and that in the ionization region by $\Delta E_I$ (see Fig. 1(b)). $\Delta E_C$ is the difference between the initial energy (5.485 MeV) of an α particle and $E_α$ at $S_1$. In Fig. 5(b), $\Delta E_C$ is plotted for various values of $D_C$. $\Delta E_P$ is the difference between $E_\alpha$ at $S_1$ and $E_\alpha$ at $S_2$, which is about 2 MeV, as plotted in Fig. 5(c) by the empty triangles for various values of $D_C$.

In Fig. 5(d), the values of $\Delta E_I$ are plotted for $D_I$ = 12, 16, 20 and 26 mm. For $D_I$ = 12 mm, $\Delta E_I$ plotted by using the squares, increases as $D_C$ increases and approaches the value of $E_\alpha$ at $S_2$ (denoted by the empty diamonds) when $D_C$ becomes 19.7 mm or longer. This means that when the sum ($D_T$) of $D_C$ (19.7 mm) and $D_I$ (12 mm) becomes 31.7 mm or longer, the α particle loses all of its kinetic energy ($E_\alpha$) at $S_2$ in the ionization region and stops there. For $D_I$ = 16 mm, the deposited energy, plotted as circles, agree with the values of $E_\alpha$ at $S_2$ if $D_C$ becomes 14.7 mm or longer. This means again that when $D_T$ = $D_C$ (14.7 mm) + $D_I$ (16 mm) = 31.7 mm or longer, all the kinetic energy ($E_\alpha$) of the α particle at $S_2$ is lost in the ionization region, and the α particle stops there. The same behavior can be observed for $D_I$ = 20 mm. For $D_I$ = 26 mm, even at the smallest value of $D_C$ = 5.7mm, $D_T$ is already 31.7 mm, and indeed the inverted triangles in Fig. 5(d) overlap with the empty diamonds, which shows that $E_\alpha$ at $S_2$ is fully deposited in the ionization region and that the length of $D_T$ = 31.7 mm is critical for the α particle to stop and lose all of its energy in the ionization region so that the ionization region can play the role of an electromagnetic calorimeter. In such a case, the calibration of our detector can be done.

In Fig. 5(e), the sum ($\Delta E_T$) of $\Delta E_C$, $\Delta E_P$, and $\Delta E_I$ is plotted as a function of $D_C$ for various values of $D_I$. For $D_I$ = 12 mm, $\Delta E_T$ agrees with the initial energy (5.485 MeV) of the α particle as denoted by the horizontal dotted line when $D_C$ = 21.7 mm, at which $D_T$ = 33.7 mm. For $D_I$ = 16 mm, $\Delta E_T$ becomes the initial energy of the α particle when $D_C$ = 17.7 mm, at which $D_T$ = 33.7 mm again. This shows that for all the values of $D_I$, $\Delta E_T$ becomes the initial energy of the α particle when $D_T$ = 33.7 mm. (Note that $\Delta E_T$ agrees with the initial energy of the α particle when $D_T$ = 33.7 mm whereas $\Delta E_I$ becomes $E_\alpha$ at $S_2$ when $D_T$ = 31.7 mm because $\Delta E_C$ and $\Delta E_P$ keep increasing as $D_C$ increases.) Thus, the energy calibration of our MICROMEGAS detector can be done for α particles by matching the measured channel number of pulse peak positions with the energy of the α particle deposited in the ionization region as long as the α particle stops and loses all of its energy ($E_\alpha$ at $S_2$) in the ionization region.

In Figs. 6 (a) and (b), Figs. 5 (d) and (e) are plotted again, but as a function of $D_T$ (= $D_C + D_I$) instead of $D_C$. Thus the curves in Fig. 5 (d) and (e) are shifted in Figs. 6 (a) and (b). For the various values of $D_C$ from 5.7 to 27.7 mm, by varying $D_I$ from 12 to 26 mm, the value of $D_T$ is changed from 17.7 to 53.7 mm. Figure 6(a) shows that the deposited energy $\Delta E_I$ becomes maximum when $D_T$ = 31.7 mm for various combinations of $D_I$ and $D_C$. If $D_T$ is less than 31.7 mm, the α particle does not stop in the ionization region and leaves the detector with part of its energy being carried away with it. If $D_T$ is larger than 31.7 mm, the α particle loses much of its energy in the collimator region before entering the ionization region and, thus, loses less energy in the ionization region, which explains the maximum peak of $\Delta E_I$ at $D_T$ = 31.7 mm.

In Fig. 6(b), the sum ($\Delta E_T = \Delta E_C + \Delta E_P + \Delta E_I$) of the energy losses are plotted for $D_I$ = 12, 16, 20, and 26 mm when $D_C$ is fixed as 19.7 mm. $\Delta E_T$ is small for small values of $D_T$ but becomes the initial energy of the α particle (5.485 MeV) at $D_T$ = 33.7 mm or longer. This shows that a value of $D_T$ of 33.7 mm is needed for an α particle to lose all of its incident energy in the volume of the detector.

For the energy calibration of our MICROMEGAS detector, we need to correlate the amount of energy deposited in the 'ionization region' with the MCA channel numbers corresponding to the pulse peak. If an alpha does not stop in the ionization region and leaves the detector with part of its incident energy carried away with it, only part of the incident energy of the alpha is deposited in the ionization region and is measured by the MCA. In that case, we cannot identify the incident energy of the alpha because the missing energy is unknown. Therefore, the length of the ionization region needs to be large enough to completely stop the alpha in the ionization region.

Figure 6(a) shows that for various combinations of $D_C$ and $D_I$, $\Delta E_I$ becomes maximum when $D_T$ = 31.7 mm. Thus, to calibrate our MICROMEGAS detector, we can associate the deposited energies ($\Delta E_I$) of the alpha in the ionization region obtained with $D_T$ = 31.7 to 53.7 mm with the channel numbers corresponding to the pulse peak positions, which will be discussed in the next subsection.

3. Experiments

We obtained the MCA spectrum of α particles for different values of $D_C$. The thickness of the ionization region is chosen as $D_I$ = 12 mm because it is around the middle of the range of $D_I$. (See Fig. 4(b).) The spectra were taken with the fixed voltages ($V_I$ = -300 V and $V_M$ = -180 V), but by changing $D_C$ from 5.7 to 31.7 mm. Figure 7(a) shows that the peak of the measured spectra moves to higher channel numbers as $D_C$ increases from 5.7 to 17.7 mm. This shows that more energy is deposited in the ionization region as $D_C$ increases. This is partly because for smaller values of $D_C$, the α particle escapes the ionization region with part of its energy being carried away with it. However, as $D_C$ increases further from 19.7 to 31.7 mm, the channel numbers for the pulse peak positions move back to lower channel numbers, as seen in Fig. 7(b). This means less energy is deposited in the ionization region as $D_C$ increases further from 19.7 mm. In other words, when the thickness of the collimator is 19.7 mm, the channel number for the pulse peak position is at its maximum. For $D_I$ = 12 mm, $D_T$ becomes 31.7 mm when $D_C$ is 19.7 mm. As already demonstrated in Figs. 5 and 6, the GEANT4 simulation shows that at $D_T$ = 31.7 mm, $\Delta E_I$ is at its maximum. Therefore, the value of $D_T$ at which the experimental channel number is at its maximum agrees with that obtained from GEANT4.

When the thickness of the collimator region increases further from 19.7 to 21.7 mm, the peak position shifts back to lower channels because the flight path of an α particle in the 'collimator' region is so large that the α particle loses much of its energy before entering the ionization region. (See Fig. 5(b), which shows $\Delta E_C$ = ~ 1.5 MeV when $D_C$ = 19.7 mm and $\Delta E_C$ = ~ 2.2 MeV when $D_C$ becomes 27.7 mm.) Thus, if $D_C$ becomes larger than 19.7 mm with $D_I$ = 12 mm, the flight path of an α particle in the 'ionization' region becomes shorter and the energy loss of an α particle in the ionization region becomes less. This experimental result is consistent with the simulation results shown in Fig. 5(d), which shows that $\Delta E_I$ is at its maximum when $D_C$ is 19.7 mm.

In Fig. 8(a), we plotted $\Delta E_I$ calculated from simulations against the channel numbers of the measured pulse peak positions for $D_I$ = 12, 16, 20, 26 mm, which correspond to $D_T = D_C + D_I$ = 17.7 to 53.7 mm. The relation between $\Delta E_I$ and the channel number shown in Fig. 8 (a) is well represented by a straight line

$$\Delta E_I \text{ (MeV)} = (5.92 \pm 0.10) \times 10^{-4} \cdot N + (0.29 \pm 0.04) \times 10^{-1}, \tag{2}$$

where N is the channel number for the pulse peak position. This relation allows us to convert the channel number of the measured pulse peak position to the energy deposited by the α particle in the ionization region or the energy of the α particle incident on the ionization region. By using this relation, we experimentally determined the energy of the α particle for various values of $D_I$ and $D_C$, and we plot the result in Fig. 8(b) by using solid symbols with dotted curves connecting them. The deposited energies calculated by using GEANT4 are also plotted in Fig. 8(b) by using open symbols with solid lines connecting them. For all the values of $D_I$, the experimentally-obtained energies of the α particle plotted by using the empty symbols agree well with the simulated deposited energies plotted by using the solid symbols.

Therefore, for the given gas mixture, the materials and the thickness of the cathode and the ionization region at room temperature and atmospheric pressure, the energy of the α particle incident on the ionization region can be determined by using Eq. (2). By adding $\Delta E_C$ and $\Delta E_P$ to the measured energy $\Delta E_I$, we can obtain the initial energy of the α particle.

4. Intrinsic Efficiency

The intrinsic efficiency of the MICROMEGAS detector for charged particles is reported as 95 ~ 99 % [1, 2, 18, 19]. To check the intrinsic efficiency of our MICROMEGAS detector for α particles, we compared our experimental results with the GEANT4 simulation results. The geometry of our MICROMEGAS detector was configured as in Fig. 1(b). While the value of $D_I$ was fixed as 20 mm, we simulated the number of α particles entering the surface $S_1$ by changing the value of $D_C$ from 5.7 to 27.7 mm. We measured the number of counts of α particles from the MCA. The intrinsic efficiency of our MICROMEGAS detector may be defined as the ratio of the measured counts of α particles to the calculated number of α particles entering the ionization region through $S_1$ for different values of $D_C$. For the present conditions of using the $^{241}$Am source (3.5 kBq) and the geometry of our MICROMEGAS detector, its average efficiency for detecting α particles is found to be (97.3 ± 2.5)%. We also estimated the number of α particles passing through the surface $S_2$. The ratio of the measured counts of α particles to the calculated number of α particles passing through the surface $S_2$ for different values of $D_C$ is (98.3 ± 1.8)%. The latter number is 1% larger than the intrinsic efficiency of 97.3% and is due to the loss of α particles in the cathode plate.

## IV. Summary

We have developed a MICROMEGAS detector to measure the energy spectrum of the α particles. Variable incident energies of the α particles from the $^{241}$Am source were obtained by changing the flight path of the α particles in the gas of the MICROMEGAS detector. The channel numbers for the measured pulse peak positions obtained by using different energies of the α particles agree well with the energy of the α particles deposited in the ionization region as calculated by GEANT4 simulations. The calibration of the MICROMEGAS detector for α particle has been done successfully. The efficiency of the MICROMEGAS detector for α particles has been found to be (97.3 ± 2.5)% under the present conditions.

## ACKNOWLEDGMENT

This work was supported in part by the Basic Science Research Program through the National Research Foundation of Korea (NRF) funded by the Ministry of Education, Science and Technology (NRF-2012R1A1A2007826, NRF-2013R1A1A2063824, NRF-2015M2B2A9032869, NRF-2015M2A2A4A01045320, NRF-2015R1C1A1A01054083.)

## REFERENCES

[1] Y. Giomataris, Ph. Rebourgeard, J. P. Robert and G. Charpak, Nucl. Instrum. Meth. Phys. Res. Sect. A **376**, 29 (1996).

[2] Y. Giomataris, Nucl. Instrum. Meth. Phys. Res. Sect. A **419**, 239 (1998).

[3] F. Sauli and A. Sharma, Annual Rev. of Nucl. Part. Sci. **49**, 341 (1999).

[4] F. Sauli, Nucl. Instrum. Meth. Phys. Res. Sect. A **477**, 1 (2002).

[5] A. Oed, Nucl. Instrum. Meth. Phys. Res. Sect. A **471**, 109 (2001).

[6] S. Andriamonje, S. Aune, G. Bignan, C. Blandin, E. Ferrer, I. Giomataris, C. Jammes and J. Pancin, Nucl. Instrum. Meth. Phys. Res. Sect. A **525**, 74 (2004).

[7] S. Andriamonje, G. Andriamonje, S. Aune, G. Ban, S. Breaud, C. Blandin, E. Ferrer, B. Geslot and A. Giganon *et al.*, Nucl. Instrum. Meth. Phys. Res. Sect. A **562**, 755 (2006).

[8] S. Andriamonje, D. Cano-Ott, A. Delbart, J. Derré, S. Diez, I. Giomataris, E. M. González-Romero, F. Jeanneau and D. Karamanis *et al.*, Nucl. Instrum. Meth. Phys. Res. Sect. A **481**, 120 (2002).

[9] J. Derré, Y. Giomataris, Ph. Rebourgeard, H. Zaccone, J. P. Perroud and G. Charpak, Nucl. Instrum. Meth. Phys. Res. Sect. A **449**, 314 (2000).

[10] J. P. Cussonneau, M. Labalme, P. Lautridou, L. Luquin, V. Metivier, A. Rahmani, V. Ramillien and T. Reposeur, Nucl. Instrum. Meth. Phys. Res. Sect. A **419**, 452 (1998).

[11] J. Derré, Y. Giomataris, H. Zaccone, A. Bay, J.-P. Perroud and F. Ronga, Nucl. Instrum. Meth. Phys. Res. Sect. A **459**, 523 (2001).

[12] J. I. Collar and Y. Giomataris, Meth. Phys. Res. Sect. A **471**, 254 (2001).

[13] J. Pancin, S. Aune, E. Berthoumieux, S. Boyer, E. Delagnes, V. Macary, B. Poumarede and H. Safa, Nucl. Instrum. Meth. Phys. Res. Sect. A **572**, 859 (2007).

[14] C. Bernet, P. Abbon, J. Ball, Y. Bedfer, E. Delagnes, A. Giganon, F. Kunne, J.-M. Le Goff and A. Magnon, Nucl. Instrum. Meth. Phys. Res. Sect. A **536**, 61 (2005).

[15] T. Alexopoulos, A. A. Altintas, M. Alviggi, M. Arik, S. A. Cetin, V. Chemyatine, E. Cheu, D. Della Volpe and M. Dris, Nucl. Instrum. Meth. Phys. Res. Sect. A **617**, 161 (2010).

[16] G. Charpak, J. Derré, Y. Giomataris and Ph. Rebourgeard, Nucl. Instrum. Meth. Phys. Res. Sect. A **478**, 26 (2002).

[17] P. K. Lightfoot, R. Hollingworth, N. J. C. Spooner and D. Tovey, Nucl. Instrum. Meth. Phys. Res. Sect. A **554**, 266 (2005).

[18] J. Bortfeldt, O. Biebel, R. Hertenberger, A. Ruschke, N. Tyler and A. Zibell, Nucl. Instrum. Meth. Phys. Res. Sect. A **718**, 406 (2013).

[19] P. Abbon, S. Andriamonje, S. Aune, T. Dafni, M. Davenport, E. Delagnes, R. de Oliveira, G. Fanourakis and E. Ferrer Ribas *et al.*, New. J. Phys. **9**, 170 (2007).

[20] D. Y. Kim, S. Andriamonje, S.-I. Bak, S. W. Hong, Y. Kadi, T.-S. Park, J. W. Shin and C. Tenreiro, Appl. Radiat. Isot. **81**, 156 (2013).

[21] R. Veenhof, Nucl. Instrum. Meth. Phys. Res. Sect. A **419**, 726 (1998).

[22] Data for the $^{241}$Am source. http://www.nucleide.org/DDEP_WG/DDEPdata.htm.

[23] P. M. Joshirao, J. W. Shin, C. K. Vyas, A. D. Kulkarni, H. Kim, T. Kim, S. W. Hong and V. K. Manchanda, Appl. Radiat. Isot. **81**, 184 (2013).

[24] P. M. Joshirao, J. W. Shin, D. Y. Kim, S. W. Hong, R. V. Kolekar and V. K. Manchanda, Nucl. Instrum. Meth. Phys. Res. Sect. B **351**, 56 (2015).

[25] S. Agostinelli, J. Allison, K. Amako, J. Apostolakis, H. Araujo, P. Arce, M. Asai, D. Axen and S. Banerjee *et al.*, Nucl. Instrum. Meth. Phys. Res. Sect. A **506**, 250 (2003).

[26] J. Allison, J. Apostolakis, H. Araujo, P. Arce Dubois, M. Asai, G. Barrand, R. Capra, S. Chauvie and R. Chytracek *et al.*, IEEE Trans. Nucl. Sci. **53**, 270 (2006).

[27] D. Attié, A. Chaus, P. Colas, E. Ferrer Ribas, J. Galán, I. Giomataris, A. Gongadze, F. J. Iguaz, R. De Oliveira *et al.*, JINST **8**, P05019 (2013).

Table 1. Ranges of the electric fields corresponding to the plateau for various lengths of the ionization region ($D_I$).

| $D_I$ (mm) | $E_I$ (kV/cm) |
|---|---|
| 4 | 0.05 - 0.35 |
| 8 | 0.10 - 0.35 |
| 12 | 0.08 - 0.35 |
| 16 | 0.14 - 0.35 |
| 20 | 0.18 - 0.35 |

Figure Captions.

Fig. 1. (a) Experimental set-up, (b) schematic diagram of the experimental set-up, and (c) schematic diagram of the MICROMEGAS detector are shown.

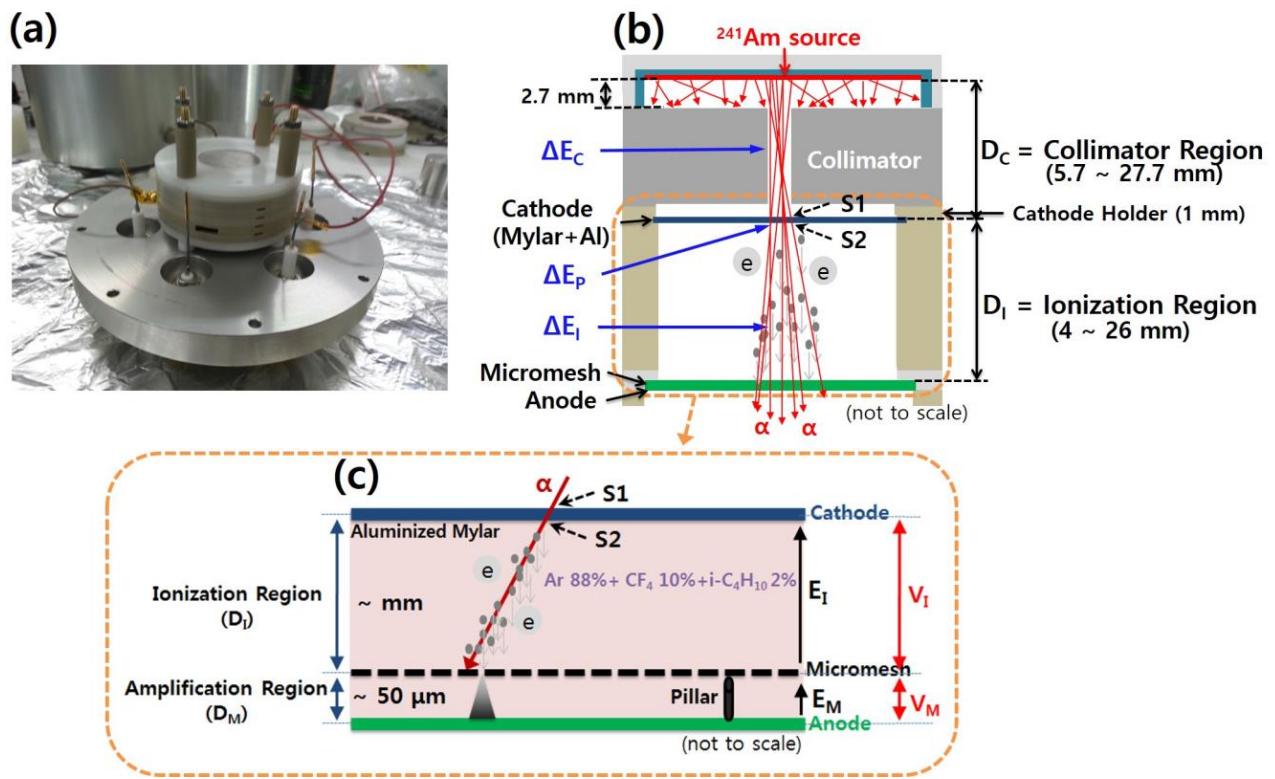

Fig. 2. Photos showing the $^{241}$Am α source (a) with and (b) without a collimator on its top.

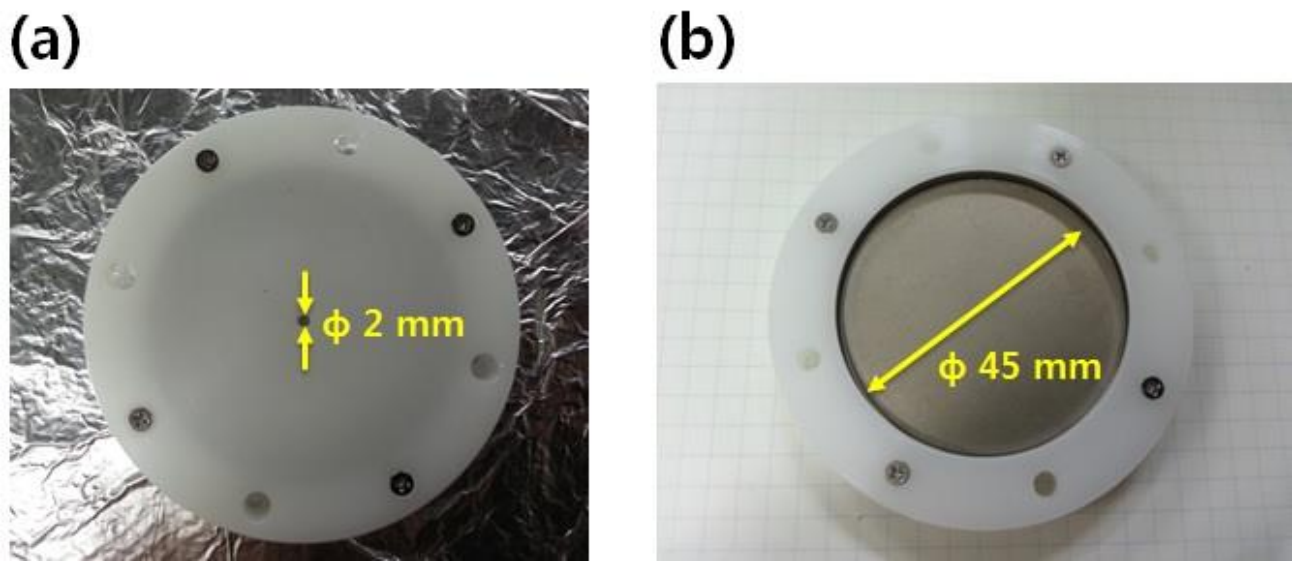

Fig. 3. Snapshot of the propagation of α particles through the collimator hole by using the GEANT4 code.

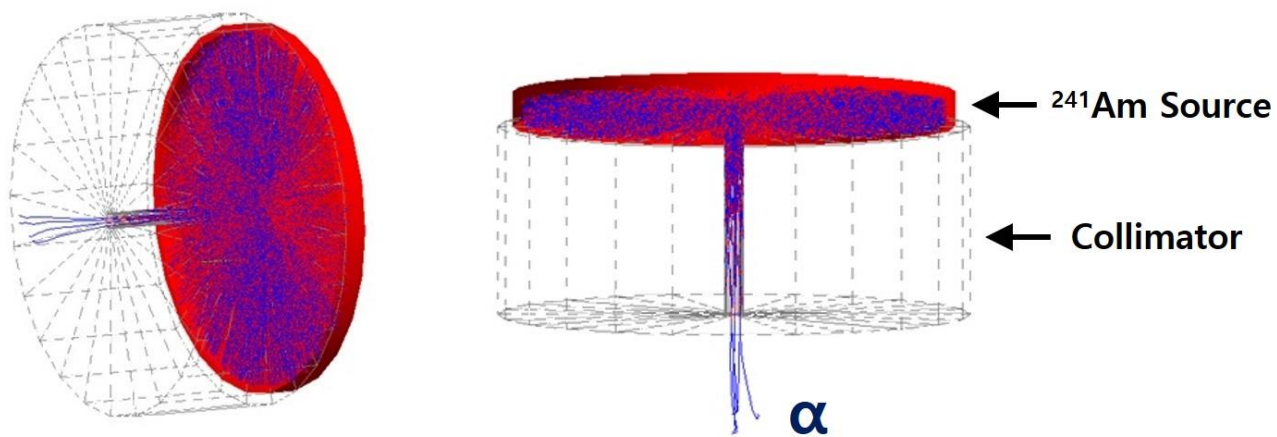

Fig. 4. (a) The output voltage obtained with α particles from the $^{241}$Am source for different values of $D_I$ when the electric field in the ionization region is fixed as $E_I = 0.25$ kV/cm. (b) The same output voltage is plotted as a function of $D_I$ for each value of $E_M$. (c) The output voltage is plotted as a function of the electric field in the ionization region ($E_I$) for different values of $D_I$ with the electric field in the amplification region fixed as $E_M = 36$ kV/cm.

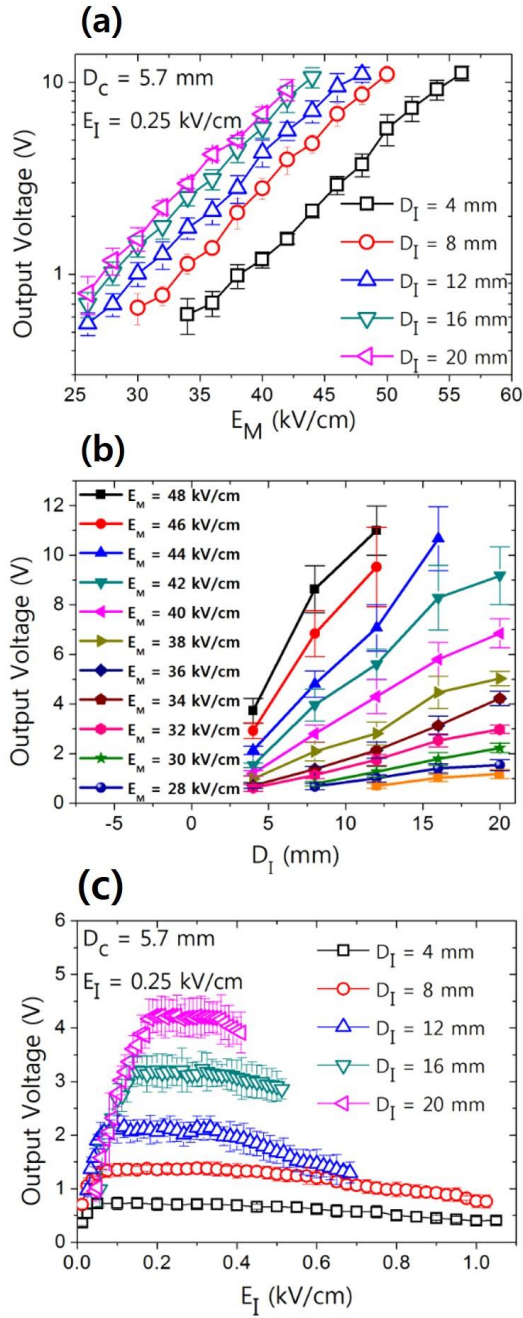

Fig. 5. (a) Calculated values of $E_\alpha$ at $S_1$ and $S_2$ are plotted as a function of $D_C$ by using the empty circles and the empty diamonds, respectively. (b) The energies deposited by the α particles in the collimator region ($\Delta E_C$), (c) the energies deposited by α particles in the aluminized mylar cathode ($\Delta E_P$), (d) the energies deposited by α particles in the ionization region ($\Delta E_I$), and (e) the sum of the deposited energies ($\Delta E_T$) are plotted for different values of $D_I$ as a function of $D_C$.

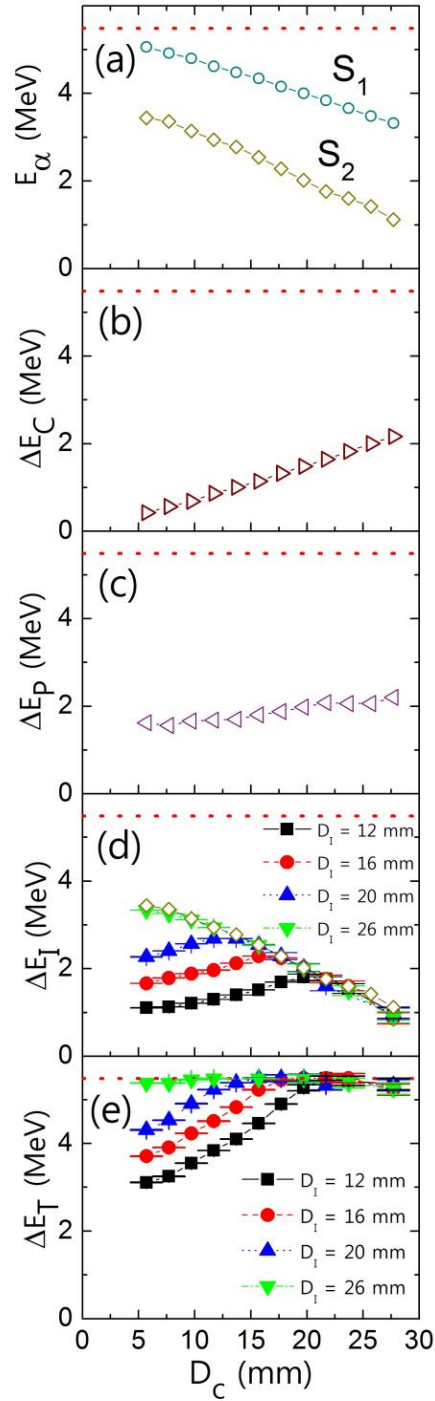

Fig. 6. (a) Energies deposited by α particles in the ionization region ($\Delta E_I$) and (b) the sum of the deposited energies ($\Delta E_T$) are plotted as functions of $D_T$ for different values of $D_I$.

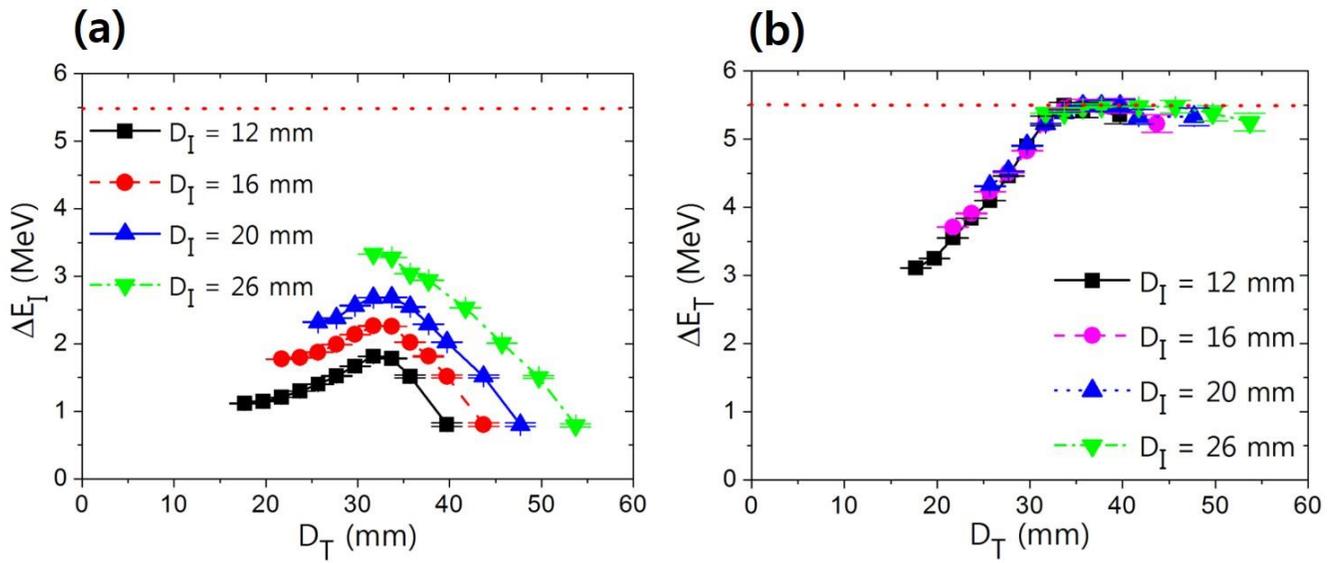

Fig. 7. MCA spectra (a) when $D_C$ increases from 5.7 to 17.7 mm (i.e., when $D_T$ increases from 17.7 to 29.7 mm) and (b) when $D_C$ increases further from 19.7 to 31.7 mm (i.e., when $D_T$ increases from 31.7 to 43.7 mm).

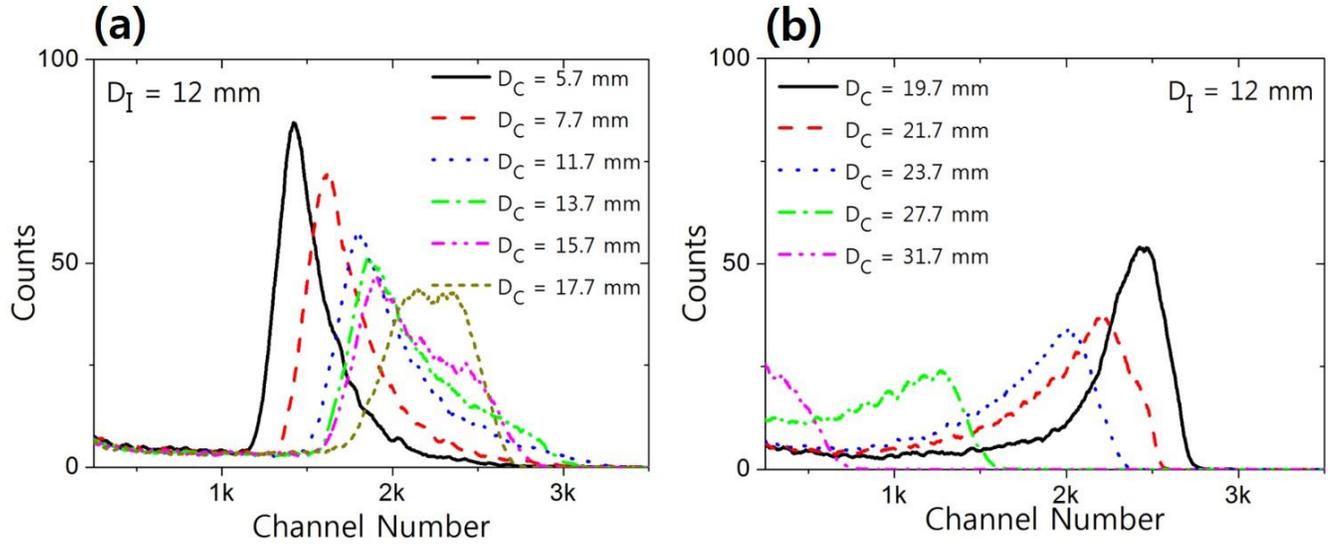

Fig. 8. (a) Relation between the deposited energies $\Delta E_I$ and the channel numbers for $D_T$ = 17.7 ~ 53.7 mm, and (b) a comparison of the converted results with the simulation results for different thicknesses of the ionization region ($D_I$ = 12 ~ 26 mm) when $D_C$ = 5.7 mm.

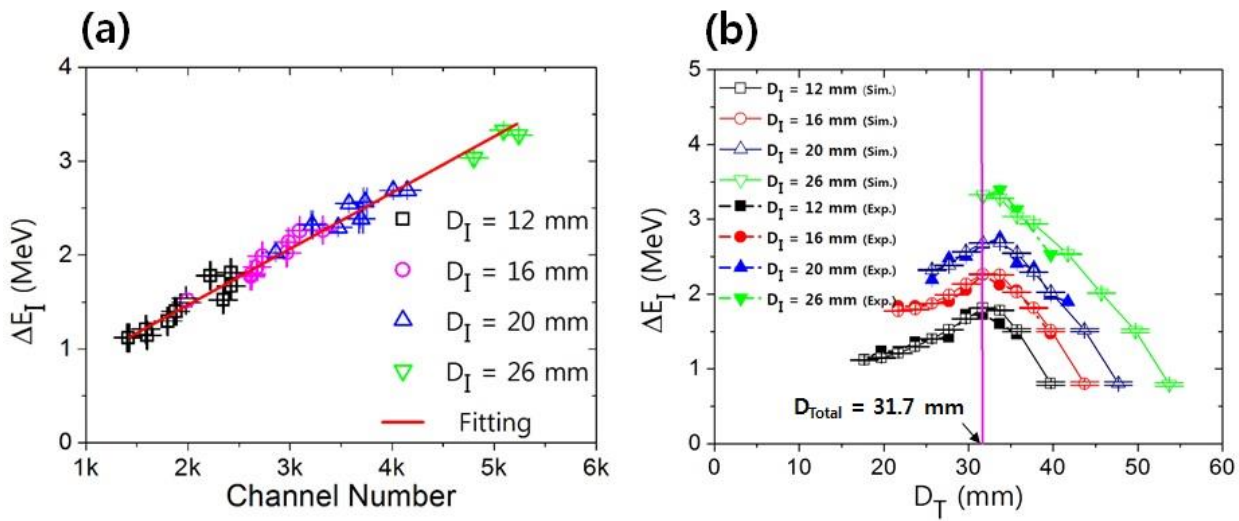